\documentclass[12pt]{article}
\usepackage{epsfig}

\newfam\mbffam \def\mbf#1{{\bf#1}}
\def\avg#1{\langle #1 \rangle}
\def\Re{\mbox{Re}}
\def\Im{\mbox{Im}}
\def\Vk{\mbf k}

\begin{document}
\rightline{UTTG-27-97}
\begin{center}
{\bf\large Effective field theory approach to Bose--Einstein
condensation}
\vskip 1cm 
{W. Vincent Liu\footnote{Electronic address: liu@physics.utexas.edu}}\\
{\it Theory Group, Department of Physics, University of Texas,
Austin, TX 78712, USA}
\end{center}
\vskip 1cm

\begin{abstract} 

We consider the low-energy collective excitations at finite
temperature of Bose--Einstein condensed gases (and liquids as well).
A most general model-independent effective Lagrangian is written down
according to a prescription obtained from the breakdown of the global
symmetry $U(1)$. To show how the theory predicts easily, we derive the
momentum and temperature dependence of the damping of excitations by
means of power counting as an example.

\flushleft{{\it PACS:} 03.75.Fi,  05.30.Jp, 11.10.Wx}

\end{abstract}
\newpage

\section{Introduction}
Bose--Einstein condensation (BEC) was achieved in 1995 in a remarkable
series of experiments on atomic vapors
\cite{M.H.Anderson++Cornell:95,Davis++Ketterle:95}, and has tremendously
boosted  both theoretical and experimental studies (for an
up-to-date review, see, e. g., Dalfovo {\it et al.} \cite{Dalfovo+:98pre}).
Among others, the study of collective excitations in BEC gives us a
great test on the finite-temperature many-body theory of interacting
Bose gases that has been
developed over last several decades. It is well known that
excitations associated with BEC 
have been well studied in the context of liquid $^4$He
\cite{Griffin:93:remark}.   
 The semi-phenomenological Landau hydrodynamic theory 
is very successful, 
but it essentially leans on an {\em ad hoc} postulated Hamiltonian
and the roles of Bose
condensation and  broken symmetry are not clear there
\cite{Khalatnikov:89}. This  theory was justified in
a sense of microscopic arguments by Feynman \cite{Feynman:72:ch11}.  On the
other hand, beginning with the work of London \cite{London:38}, 
a microscopic theory of superfluid $^4$He using field theoretical methods
has also been 
extensively investigated through a weakly interacting gas model
\cite{Griffin:93:remark}.  The most important results obtained from
this model  are
based on the presence of a Bose condensate, which breaks down the global
symmetry $U(1)$ of the system. 
In fact, 
the concept of broken symmetry originated in particle
physics has been brought into the low temperature phenomena of
superfluidity and 
superconductivity first  by Anderson \cite{Anderson:84}, and now it is
believed that it
controls much of dynamics of many-body systems.

In the microscopic study of  a Bose condensed gas,
one usually needs a detailed model like that of Bogoliubov, Beliaev 
\cite{Beliaev:58}, or Gross-Pitaevskii
to explain the mechanism for a 
 Bose condensate, and as a basis  for approximate
quantitative calculations, but not to derive the most important exact
consequences of the spontaneous symmetry breakdown.
(Of course, such a microscopic model itself  is an approximation.) 
Yet, if the general features are in fact model independent consequences of
the spontaneous breakdown of the $U(1)$ symmetry, 
why can't we derive
them directly from this breakdown by finding the most  general
effective Lagrangian of the system?
An answer to this question is provided by
a standard technique of modern quantum field theory
\cite{Weinberg:96:ch19+21}, called the effective field theory approach
to the symmetry breaking.
It was applied to superconductivity \cite{Weinberg:86,Weinberg:94} where 
the fundamental properties of conventional superconductors
such as Meissner effect and flux quantization
are derived directly from the spontaneous breakdown of
electromagnetic gauge 
invariance  without using a detailed model like that of
Bardeen-Cooper-Schrieffer (BCS)  theory.

This paper aims to show that many important properties of low-energy 
excitations in a generic
homogeneous Bose system  can be essentially determined as
the exact consequence of spontaneously broken symmetry without introducing
unnecessary approximations. Assuming the 
$U(1)$ invariance is spontaneously broken, we shall derive the most general
effective Lagrangian of the
system by performing procedures very much analogous to that of
pion-nucleon effective field theory
\cite{Weinberg:remark} in quantum
chromodynamics. The present paper gives an alternative derivation of the
Popov effective Lagrangian that we 
used for calculating the damping rate of the excitation
\cite{Liu:97b,damping:remark}.  As we
shall see, the whole procedure is incredibly simple and straightforward, but
also gives a feasible way of predicting some experimental observables.
An example on the damping rate  shall be given 
to show how the theory works easily.
The purpose of the present work is twofold. First, it offers a deep
way of justifying the Landau
theory of quantum hydrodynamics and explaining why a weakly
interacting gas model 
can yield some meaningful results for strongly interacting 
superfluid $^4$He. Second, it shows a simple, powerful effective field
theory approach to investigating the  excitation dynamics of a
homogeneous Bose 
condensed system.  

\section{Effective Lagrangian}

We start by considering a simple nonrelativistic many-body problem of
spinless bosonic particles  at finite temperature. 
In units of $\hbar\equiv 1$ and $k_B\equiv1$,
its Euclidean action functional is given by~\cite{Popov:83,Negele+:88}
\begin{equation}
I[\psi,\psi^\dag] =\int_0^\beta d \tau \int_{-\infty}^\infty d^3 x
{\cal L} \label{eq:I}
\end{equation}
with the Lagrangian density
\begin{eqnarray}
{\cal L} & =& 
{1\over 2} \left [ \psi^\dag(x) \partial_\tau \psi(x)- \psi(x)
\partial_\tau \psi^\dag(x) \right] -
{1 \over 2m}\nabla\psi^\dag(x) \cdot
\nabla\psi(x)  -\lambda \left[\psi(x)^\dag\psi(x)\right]^2  \label{eq:L}
\end{eqnarray}
where  $\beta=1/T$ denotes the inverse of
temperature.
Here, we  write
$x=(\mathbf{x},\tau)$
and $\tau$ is the imaginary ``time''.
All fields fulfill the boundary conditions as of 
$\psi(\mathbf{x},\tau+\beta) =\psi(\mathbf{x},\tau)$.
Obviously, the action (\ref{eq:I}) is Galilean-invariant, and 
 is invariant under the global phase
transformations of group $U(1)$,
$$\psi(x) \rightarrow \exp(i\Lambda) \psi(x),$$
with $\Lambda$ an arbitrary constant.
This symmetry is known to be completely broken below
some critical temperature $T_c$, i.e., a
Bose--Einstein
condensate appears with $\avg{\psi} \neq 0$ \cite{Forster:75:ch10}. 
The particle density and current associated with the symmetry are
${\mbf J} = (i/ 2m)(\psi^\dag\nabla\psi -
\psi\nabla\psi^\dag)$,
$J_4 =i\psi^\dag\psi \equiv i \rho$,
with the current conservation
$
i\partial_\tau \rho +\nabla \cdot {\mbf J}=0.
$

According to our
general understanding of spontaneously broken symmetries
\cite{Weinberg:96:ch19+21}, 
any system described by an action with symmetry group $G$, when in
a phase in which $G$ is spontaneously broken to a subgroup  $H$, will
possess a set of Goldstone modes, described by fields that transform
under $G$ like the coordinates of the coset space $G/H$. 
In our case, there will be a single Goldstone mode described by a
(massless) real scalar field
$\phi(x)$ that transforms under $G=U(1)$ like the phase
$\Lambda$ itself. The group $U(1)$ has the multiplication rule
$g(\Lambda_1)g(\Lambda_2) = g(\Lambda_1+\Lambda_2)$, so under a phase
transformation with parameter $\Lambda$ the field
$\phi(x)$ will undergo the transformation
\begin{equation}
\phi(x) \rightarrow \phi(x) +
\Lambda. \label{eq:phi_transform} 
\end{equation}

In low temperature physics, the Goldstone mode is accompanied with
another excitation, known as the density (order parameter)
fluctuation, which we will 
see to have nearly zero
frequency in the long-wavelength limit. 
Both together form a non-trivial irreducible linear representation of
the group $U(1)$. 
To see the theory must involve the Goldstone fields, we may write all
ordinary complex fields as \cite{Popov:72}
\begin{equation}
\psi(x) =\sqrt{\rho(x)} \exp(i\phi(x))
 \label{eq:psi}
\end{equation}
where  the $\rho$ is  the density field. 
Under phase transformation, the $\rho$ is invariant while $\phi$
transforms according to the rule (\ref{eq:phi_transform}). 
Now,  rewritten in terms of fields $\phi$
and $\rho$, the Lagrangian (\ref{eq:L})   becomes 
\begin{eqnarray}
{\cal L} & =& \displaystyle
 	\rho i\partial_\tau\phi
	-{(\nabla\rho)^2 \over 8m\rho}  
	-{\rho (\nabla\phi)^2 \over 2m}-\lambda \rho^2
 	. \label{eq:L2}
\end{eqnarray}
If the $U(1)$ symmetry is broken, we have $\rho_0 \equiv|\avg{\psi}_0|^2
 \neq 0$ where
$\avg{\cdots}_0$ indicates the expectation value over the ground state.
It follows that we may write $\rho(x) =\rho_0 +\sigma(x)$, where the
$\sigma$ field describes density fluctuations, known also as
collective modes. The effective
Lagrangian for $\phi$ and $\sigma$ then reads
\begin{eqnarray}
{\cal L}_{\mbox{eff}} & =& \displaystyle
	\sigma i\partial_\tau\phi
	-{(\nabla\sigma)^2 \over 8m(\rho_0+\sigma)}  
	-{(\rho_0+\sigma) (\nabla\phi)^2 \over 2m} -\lambda 
	(\rho_0+\sigma)^2  , \label{eq:Leff}
\end{eqnarray}
where we have ignored all total derivatives.
For later use, we record the current conservation in terms of $\phi$
and $\sigma$ explicitly:
\begin{equation}
i\partial_\tau \sigma -{\nabla \cdot [(\rho_0+\sigma)\nabla\phi] \over
m}=0. 	\label{eq:Jconservation2}
\end{equation}

Here is an important point: to derive Eq.~(\ref{eq:Leff}) it was not
really necessary to start with the model Lagrangian (\ref{eq:L}). Indeed,
according to our understanding of spontaneously broken symmetry, we
did not need to start with {\it any} specific theory. A familiar
example in particle physics is the effective theory of pion-nucleon
interaction with $SU(2)\times SU(2)$ spontaneously broken to $SU(2)$.
The important
thing is that Eq.~(\ref{eq:Leff}) is invariant under the $U(1)$
transformation. 
The general  
theory of broken symmetries (see, for example, Weinberg
\cite{Weinberg:96:ch19+21}) tells us that, for
symmetry group $U(1)$,  the 
most general form of the effective Lagrangian for the density
fluctuation field and the
Goldstone field must be constructed solely from the ingredients
$\sigma$, $\nabla\sigma$,
$\partial_\tau \sigma$, $\nabla\phi$ and $\partial_\tau \phi$
together with higher derivatives of these objects.
There are also two additional rules that the Lagrangian must obey. 
The first rule is the Galilean invariance for a
non-relativistic system such that the combination of $i \partial_\tau \phi -
(\nabla\phi)^2 /2m$  must always appear 
together in the effective Lagrangian and likewise 
$i \partial_\tau \sigma -(1/m)\nabla\phi \cdot
\nabla\sigma$~\cite{Galilean:remark}. 
This has been understood through
investigations by
Takahashi \cite{Takahashi:88} and  Greiter {\it et al.}
\cite{Greiter+:89}.
The second rule is the  time-reversal 
symmetry such that the action is
invariant under the transformation
$$\phi \rightarrow -\phi, \quad
\tau \rightarrow -\tau. 
$$
This is observed from the action (\ref{eq:I}).
This symmetry requires that only the even powers of the
Galilean invariant $[i\partial_\tau\sigma -(1/m)\nabla\phi \cdot \nabla\sigma]$
be included in the Lagrangian. For instance, the possible lowest power of it
is $[i\partial_\tau\sigma -(1/m)\nabla\phi \cdot \nabla\sigma]^2$.
This is equivalent to have terms of
$[(\rho_0+\sigma)\nabla^2\phi/m]^2$ by making use of the current
conservation (\ref{eq:Jconservation2}) \cite{SigmaTerm:remark}.

Hence,  according to the above prescription 
the most general $U(1)$-invariant action functional takes the
following form
\begin{equation}
I_{\mbox{eff}}[\sigma,\phi] =
\int_0^\beta d \tau \int_{-\infty}^\infty d^3 x {\cal L}_{\mbox{eff}}
\label{eq:Ieff}
\end{equation}
with
\begin{eqnarray}
{\cal L}_{\mbox{eff}} & =& \displaystyle
 	\sigma \left[i\partial_\tau\phi -{(\nabla\phi)^2
 \over 2m}\right] -{F_\phi\over 2m} (\nabla\phi)^2 
	-{F^\prime_\phi m\over 2}\left[i\partial_\tau\phi -{(\nabla\phi)^2
 \over 2m}\right]^2
\nonumber \\
 && -{F^{\prime\prime}_\phi m^2\over 3!}\left[i\partial_\tau\phi
 	-{(\nabla\phi)^2  \over 2m}\right]^3
   -{F_\sigma \over 2m} (\nabla\sigma)^2
   -{c_2\over 2m}\sigma^2 - {c_3\over 3!m }\sigma^3 
\nonumber \\
&& 	- {c_3^\prime m\over 2} \sigma 
	\left[i\partial_\tau\phi -{(\nabla\phi)^2\over 2m}\right]^2
	- {c_3^{\prime\prime} \over 2} \sigma^2 
	\left[i\partial_\tau\phi -{(\nabla\phi)^2\over 2m}\right]
+\cdots . \label{eq:Leff2}
\end{eqnarray}
The terms indicated by $\cdots$ will contain higher powers and/or
derivatives  of the
$\sigma$ and/or  $\phi$ fields.
Any term of a total derivative has been ignored.
The coefficients
$F_\phi,F^\prime_\phi,F^{\prime\prime}_\phi,F_\sigma,c_2,c_3,c_3^\prime$
and $c_3^{\prime\prime}$ 
have the dimensions of
$K^{3},K,K^{-1},K^{-3}, K^{-1},K^{-4},K^{-2}$ and $ K^{-3}$, 
respectively, where $K$
represents a typical 
momentum scale that shall be discussed in the next section.
(Here,  we adopt a normalization such that
$\phi$ is dimensionless and $\sigma$ has the dimension of
$K^3$.) Similar
effective Lagrangians appeared in 
different contexts, but were all based on some
microscopic model.  For
instance, Popov \cite{Popov:72} derived it by means of the power
expansion of the 
pressure (or equivalently grand potential) in terms of inhomogeneity 
for Bose gases, 
Aitchison {\it et al.} \cite{Aitchison+:95} found it equivalent to a
time-dependent non-linear Schr\"odinger Lagrangian for BCS
superconductors at $T=0$, and Demircan {\it et al.} \cite{Demircan+:96}
implied that it could be obtained from the Feynman wave function
of superfluids. 
If $-(1/m)\nabla\phi$ and $\sigma$ are identified with the phonon
velocity field $\mbf v$ and the  density variation $\rho^\prime$ 
of Ref.~\cite{Khalatnikov:89}, respectively, 
one finds the action (\ref{eq:Ieff}) corresponds to a
Hamiltonian
\begin{equation}
\int d^3 x \left\{ 
{F_\phi\over 2} m v^2+ {F_\sigma \over 2m} (\nabla\rho^\prime)^2 +
{c_2\over 2m}{\rho^\prime}^2 + {c_3\over 3!m}{\rho^\prime}^3  
+{1\over 2} m \rho^\prime v^2 + \cdots \ \right\},
\end{equation}
which is one form of the Landau-Khalatnikov hydrodynamic Hamiltonian
\cite{Khalatnikov:89:eq7-12}. Hence, the Landau quantum
hydrodynamics is the
exact consequence of  the breakdown of the $U(1)$ symmetry.

\section{Power Counting}

Many-body properties can be studied through 
propagators (or Green's functions).
In the following, we write $\phi$ and $\sigma$ fields into a real $2$-component
scalar  
$$\Phi = {\phi \choose \sigma}$$
with Greek indices $\alpha,\beta,\cdots$ ($=1,2$) labeling its components.
The propagators are defined by the matrix
$$ \displaystyle
\Delta_{\alpha\beta}(x-x^\prime) =\avg{T\{\Phi_\alpha(x)\Phi_\beta(x^\prime)\}} =
{\int [\prod_x d\phi(x)d\sigma(x)] \Phi_\alpha(x)\Phi_\beta(x^\prime)
\exp{I_{\mbox{eff}}[\phi,\sigma]} \over \int [\prod_x d\phi(x)d\sigma(x)]
\exp{I_{\mbox{eff}}[\phi,\sigma]}}, $$
where $T$ denotes a time-ordered product on $\tau,\tau^\prime$.
Consider the quadratic part of the
effective action
\begin{eqnarray}
I_{\mbox{eff}}^{\mbox{quad}} &=& 
	\displaystyle\int_0^\beta d \tau \int_{-\infty}^\infty d^3 x
 	\left[ \sigma i\partial_\tau\phi -{F_\phi\over 2m}
	(\nabla\phi)^2  + {F^\prime_\phi m\over 2} (\partial_\tau\phi)^2
	-{F_\sigma \over 2m} (\nabla\sigma)^2
 	-{c_2\over 2m}\sigma^2 
\right] \nonumber \\
 &\equiv &- {1\over 2} 
\int d^4x d^4x^\prime \Phi^\dag(x) {\cal D}(x,x^\prime) \Phi(x^\prime),
	\label{eq:I^quad}
\end{eqnarray}
where
\begin{equation}
{\cal D}(x,x^\prime) = \left(
\begin{array}{cc}
[-{F_\phi\over m}\nabla_x^2 +F^\prime_\phi m \partial^2_\tau]
 \delta^4(x-x^\prime) & 
i\partial_\tau \delta^4(x-x^\prime) \\
-i\partial_\tau \delta^4(x-x^\prime) & 
[-{F_\sigma\over m} \nabla_x^2
+ {c_2\over m}]\delta^4(x-x^\prime) 
\end{array}  \label{eq:D(x)}
\right).
\end{equation}
The  free propagators are given by the inverse of the matrix ${\cal
D}$:
\begin{equation}
\Delta(x,x^\prime)= {\cal D}^{-1}(x,x^\prime).
\end{equation}
The calculation of propagators is simplified by transforming to
momentum basis via the following Fourier transformation
\begin{equation}
\Delta(x,x^\prime)= {1 \over \beta (2\pi)^3} \sum_\nu \int d^3 k
\Delta(k) e^{i{\Vk} \cdot ({\mbf x}-{\mbf x}^\prime) -
i\omega_\nu (\tau-\tau^\prime)},  \qquad
\label{eq:Fourier}
\end{equation}
where Matsubara frequencies $\omega_\nu\equiv 2\pi \nu/\beta$ 
($ \nu=0,\pm 1,\pm 2,\cdots$) and 
the $4$-momentum notation $k=(\Vk,\omega_\nu)$ is used.
We then have 
\begin{equation}
\Delta^{-1}(k)={\cal D}(k) = \left(
\begin{array}{cc}
{F_\phi\Vk^2 / m} -F^\prime_\phi m \omega_\nu^2 &  \omega_\nu \\
-\omega_\nu  & (F_\sigma \Vk^2 +c_2)/m
\end{array} 
\right). 	\label{eq:D(k)}
\end{equation}
By finding its inverse matrix, the free propagators (see
Fig.~\ref{fig:rules}) are  
\begin{equation}
\Delta(k)= {[1-F^\prime_\phi(F_\sigma \Vk^2+c_2)]^{-1} 
\over\omega_\nu^2 +\epsilon^2(\Vk) } \left( 
\begin{array}{cc}
(F_\sigma \Vk^2 + c_2)/m &
-\omega_\nu  \\
\omega_\nu   &
F_\phi\Vk^2/m -F^\prime_\phi m \omega_\nu^2
\end{array} 
\right)	\label{eq:Delta(k)}
\end{equation}
with the energy spectrum 
\begin{equation}
\epsilon (\Vk) = {1\over m}\sqrt{F_\phi\Vk^2(F_\sigma\Vk^2 +
c_2)  \over  1-F^\prime_\phi(F_\sigma \Vk^2+c_2)} \ . \label{eq:spectrum} 
\end{equation}
We shall consider the low momentum-energy region such that 
$k \equiv |\Vk| \ll k_0\equiv \sqrt{c_2/F_\sigma}$, in which the
spectrum reduces to the phonon type 
$$
\epsilon(\Vk) \simeq ck
$$
with the phonon velocity $c\equiv (1/m)\sqrt{F_\phi
c_2/(1-F^\prime_\phi c_2)}$. Notice that
the energy spectrum is linear in $k$, vanishing as $k\rightarrow 0$.

In the calculation of Feynman diagrams at finite temperatures in Euclidean
field theory, one encounters the summation over discrete Matsubara
frequencies. A standard technique (see, for example, \cite{Kadanoff+:62})
is available to perform such a
Matsubara summation. The trick is to use a contour integral in the
complex energy plane.  Let $h(\omega)$ be a function of
complex variable $\omega$, analytical on the line Re$\omega=0$, which
decreases faster than $1/|\omega|$ as $|\omega| \rightarrow \infty$.
 We then have 
\begin{equation}
{1\over \beta}\sum_\nu h(i\omega_\nu) = -\oint_C {d\omega\over 2\pi i}
f(\omega) h(\omega) \ ,
\label{eq:sum-int}
\end{equation}
where
$$ f(\omega) = {1\over e^{\beta \omega} -1},$$
and the $C$ is a contour in complex $\omega$-plane encircling all poles
of  function $h(\omega)$ in a positive sense (but those of
function $f(\omega)$ in a negative sense).   

Now consider a general process involving arbitrary numbers of the
Goldstone field $\phi$ and the density fluctuation field $\sigma$. We
suppose that their energies and momenta and the thermal energy ($\sim T$)
are all at most of some order 
$K$, which is small compared with $k_0$ defined above.
Even though Lagrangians like (\ref{eq:Leff2}) are not renormalizable in
the usual sense, we saw in, for example, 
the pion-nucleon theory \cite{Weinberg:96:ch19+21}
that such Lagrangians can yield finite results as long as they contain
all possible terms allowed by symmetries, for then there will be a
counterterm available to cancel every infinity. 
If we define the renormalized values of the constants
$F_\phi,F^\prime_\phi, F^{\prime\prime}_\phi,F_\sigma,c_2,c_3, c_3^\prime,
\cdots$ in ${\cal L}_{\mbox{eff}}$ 
by specifying the values at energies of order
$K$, then the integrals in momentum-space Feynman diagrams will be
dominated by contributions from virtual momenta which are also of
order $K$ (because renormalization makes them finite, and there is no
other possible effective cut-off in the theory). We can then develop
perturbation theory as a power series expansion in $K$. 

Each derivative in each interaction vertex contributes one factor of
$K$ to the order of magnitude of the diagram; internal propagators
$\Delta_{11} (k)$, $ \Delta_{12} (k)$
$(=-\Delta_{21} (k))$, and $\Delta_{22} (k)$
contribute factors of $K^{-2}$, $K^{-1}$ and a unit, respectively; and
each integration volume $d^4 k$ ($\equiv d^3 k d\omega$) 
associated with the loops of the
diagram contributes a factor of $K^4$. So a general connected diagram
make a contribution of order $K^\nu$, where
\begin{equation}
\nu= -2 I_\phi - I_\times + \sum_i d_i V_i + 4L. \label{eq:nu}
\end{equation}
Here $d_i$ is the number of derivatives in an interaction of type $i$,
$V_i$ is the number of interaction vertices of type $i$ in the
diagram, $I_\phi$ and $I_\times$ are
the numbers of internal $\phi$ lines and $\phi$-$\sigma$ cross lines,
respectively, and $L$ is the number of loops. 
There is a familiar topological relation for connected graphs:
\begin{equation}
2I_\phi + I_\times  +E_\phi= \sum_i \phi_i V_i,
\end{equation}
where $\phi_i$ is the number of Goldstone fields $\phi$ in
interactions of
type $i$ and $E_\phi$ is the number of external $\phi$ lines.
Eliminating the quantity $I_\phi$, the two topological equations above give
\begin{equation}
\nu= \sum_i (d_i-\phi_i)V_i + E_\phi + 4L. \label{eq:nu2}
\end{equation}
The important point here is that the coefficient $d_i-\phi_i$ in the
first term is always positive or zero. 
Hence, with the numbers of loops and external $\phi$-lines fixed, 
the leading terms are those graphs of $d_i-\phi_i=0$. The
interactions that satisfy this condition are of no derivatives of the
$\sigma$ field, noticing that the Goldstone fields always appear with
derivatives. Next, we shall see the power counting (\ref{eq:nu2}) provides us
a tool that ensures that only finite number of terms of ${\cal
L}_{\mbox{eff}}$ are necessary.   

\section{Low-energy Excitation Spectrum}

\begin{figure}[tbp]
\begin{center}
\epsfig{file=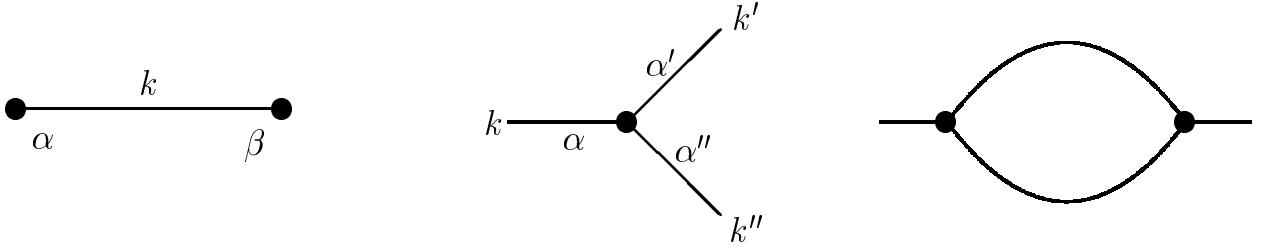,width=\linewidth}
\end{center}
\caption{\small Feynman diagrams of propagators, vertices, and the
one-loop selfenergy
$\Pi$. Notations are $k=(\Vk,\omega_\nu)$ and
$\Phi_\alpha^\dag=(\phi,\sigma)$ with Greek indices $\alpha, \beta, \cdots=1,2$.}
\label{fig:rules}
\end{figure}

Let us now apply this method to the calculation of higher order
corrections to the propagator,
$$\Delta^\prime(k)=\Delta(k) + \Delta(k)\Pi(k) \Delta^\prime(k)$$
where the matrix $\Pi(k)$ denotes  the self-energy connected
graphs. The lowest order correction shall be from the graphs of
one loop with $L=1$ in Eq.~(\ref{eq:nu2}), to which only cubic
interactions of the Lagrangian (\ref{eq:Leff2}) contribute.  
The following are all cubic interactions that concern us:
$$ -{\sigma(\nabla\phi)^2 \over 2m},\ {F^\prime_\phi \over 2}
i\partial_\tau \phi (\nabla\phi)^2, 
\ -{F^{\prime\prime}_\phi m^2 \over 3!} (i\partial_\tau\phi)^3,
  \
-{c_3\over 3!m }\sigma^3, \ 
{c_3^\prime m\over 2} \sigma(\partial_\tau\phi)^2, \ \mbox{and} \
-{c_3^{\prime\prime} \over 2}\sigma^2 i\partial_\tau\phi,
$$
which give the following vertices
\begin{equation}
\begin{array}{c}
 \delta^4(k+k^\prime+k^{\prime\prime})
\left\{ \displaystyle
\sum_{P\{ { 
\alpha\rightarrow \alpha^\prime \rightarrow \alpha^{\prime\prime}
\rightarrow \alpha \atop
k\rightarrow k^\prime \rightarrow k^{\prime\prime} \rightarrow k
} \}}  \left[
\delta_{\alpha,1} \delta_{\alpha^\prime,1} 
\delta_{\alpha^{\prime\prime},2} \left( {(\Vk\cdot\Vk^\prime)\over m} 
-c_3^\prime m\omega_\nu\omega_{\nu^\prime} \right)
 -\delta_{\alpha,1} \delta_{\alpha^\prime,2} 
\delta_{\alpha^{\prime\prime},2} c_3^{\prime\prime} \omega_\nu \right]
\right. 
 \\
\displaystyle \left.
-\delta_{\alpha,1} \delta_{\alpha^\prime,1}
\delta_{\alpha^{\prime\prime},1}\left[  
\sum_{P\{k\rightarrow k^\prime \rightarrow
k^{\prime\prime} \rightarrow k \}}  F^\prime_\phi 
(\Vk\cdot\Vk^\prime)\omega_{\nu^{\prime\prime}}
+ F^{\prime\prime}_\phi m^2 \omega_\nu\omega_{\nu^\prime}
\omega_{\nu^{\prime\prime}} 
 \right] 
-\delta_{\alpha,2}
\delta_{\alpha^\prime,2} \delta_{\alpha^{\prime\prime},2} {c_3\over m}
 \right\} ,
\end{array}   \label{eq:vertex}
\end{equation}
where we define $\delta^4(k+k^\prime+k^{\prime\prime}) =
\delta^3(\Vk+\Vk^\prime+\Vk^{\prime\prime}) \delta_{\nu+ \nu^\prime 
+\nu^{\prime\prime},0}\ $, and $\sum_{P\{\cdots\}}$ indicates the sum
over three cyclic permutations. 
(The vertices and the one-loop diagram of $\Pi$ are depicted in
Fig.~\ref{fig:rules}.)  
In the following, we will show how some fundamental low temperature
properties of low energy excitations can be easily derived without
complicated calculations.

The spectrum of excitations is given by the poles of the
exact propagators (Green's functions)  that are
solutions to the following equation
\begin{equation}
\det {\Delta^\prime}^{-1}(k) = \det(\Delta^{-1}(k) -\Pi(k)) =0.
\label{eq:det}
\end{equation}
From this expression, the spectrum can be obtained by analytically
continuing it to values of 
$\omega_\nu$ that are not Matsubara frequencies by doing $i \omega_\nu
\rightarrow \omega 
+i\eta$ ($\eta\equiv 0^+$), but of course this shall only be done
after the Matsubara frequency sum.  From the free propagator
expression (\ref{eq:Delta(k)}), the (real) leading term of the
spectrum is obviously given by $\omega=\epsilon(\Vk)$,  and  
we shall approximate
$\Re\omega=\epsilon(\Vk)$ as far as only the lowest corrections are concerned. 
Thus the  spectrum is given in the following form  
\begin{equation}
\omega =\epsilon(\Vk)-i\gamma(\Vk). \label{eq:spectrum2}
\end{equation}
Now our task is
to find the leading order of the imaginary part of the spectrum,
called the damping rate $\gamma$.   
After some elementary algebras, keeping only leading order in $k$, 
Eqs.~(\ref{eq:D(k)}) and (\ref{eq:det}) yield 
\begin{eqnarray}
\displaystyle
\gamma(\Vk) &=&\left\{ 
{c_2 \over 2 m \epsilon(\Vk)} \Im\Pi_{11}(\Vk, -i\omega+\eta)
	+ {F_\phi k^2 + F^\prime_\phi m^2\epsilon^2(\Vk) 
\over 2 m\epsilon(\Vk)} \Im\Pi_{22}(\Vk,-i\omega+\eta) \right. \nonumber 
\\
  &&\left. -{1\over 2} \Re\left[\Pi_{12}(\Vk, -i\omega+\eta)-
			\Pi_{21}(\Vk, -i\omega+\eta)\right] 
\right\}(1-F^\prime_\phi c_2)^{-1}
\label{eq:gamma}
\end{eqnarray} 
with $\Pi_{12}(k) = -\Pi_{21}(k)$.
One can verify that Eq.~(\ref{eq:gamma}) can reduce to the specific
forms previously presented in Refs.~\cite{Popov:72,Liu:97b}.
Rather than go on to really calculate all elements of the self-energy
matrix, we shall try to quickly derive the temperature and
momentum dependences of the damping  by using the power counting
formula (\ref{eq:nu2}). 

At $T=0$,  the typical momentum scale $K$ is just the momentum $k$ of a
excitation that is carried into $\Pi$ graphs through external lines. 
From the expression (\ref{eq:nu2}), we find 
$$
 \Pi_{11}(k) \sim k^6, \quad \Pi_{12}(k) = -\Pi_{21}(k) \sim k^5, \quad
\mbox{and} \quad \Pi_{22}(k) \sim k^4, 
$$
so that Eq.~(\ref{eq:gamma}) immediately yields
\begin{equation}
\gamma_{T=0}(\Vk) \propto k^5. \label{eq:gammaT=0}
\end{equation} 
We therefore have  determined
$\gamma(\Vk)$ up 
to some coefficient that only depends on something  other than $k$ (e.g., the 
interaction strength).
Expression (\ref{eq:gammaT=0})
agrees with the well-known result first derived by
Beliaev in 1958 \cite{Beliaev:58} for a weakly interacting dilute gas model by
calculating the Green's function till the second order approximation.
However, our result has been derived without assuming a weak coupling,
so it is valid for general Bose condensed liquids including superfluid $^4$He.

For temperature such that  $ck \ll T\ll ck_0$, 
we can follow the same arguments 
as we had for $T=0$, but  some additional
examination on diagrams is demanded in order to get the correct $k$ and $T$
dependences of $\gamma$.  Each self-energy
element $\Pi_{\alpha\beta}$ carrying a momentum $k$
yields a contribution of order $K^\nu$,
which for $T\neq 0$ can be decomposed into $K^\nu=k^lT^{\nu-l}$
($l=0,1,\cdots, \nu$). For temperature regime concerned here, the
$k$-dependence of the   $\Pi$  arises only from those
vertices that are associated with external (input) momenta  after taking
off the $\delta$-function dependence from the vertex expression
(\ref{eq:vertex}).
Hence, $k^l$ can be determined by simply  counting the power of external
(input) momenta from each vertex.  
In other words, $l$ is equal to the number of external $\phi$ lines
attached to each $\Pi$ graph.
Thus, 
$$ \Pi_{11}(k) \sim k^2 T^4, \quad \Pi_{12}(k) = -\Pi_{21}(k) \sim k
 T^4, \quad 
 \mbox{and} \quad \Pi_{22}(k) \sim T^4, 
$$
in comparison with those of $T=0$.
From Eq.~(\ref{eq:gamma}), we  find the damping rate satisfying 
\begin{equation}
\gamma(\Vk) \propto k T^4. \label{eq:gammaT}
\end{equation}
Again, in the context of a weakly interacting dilute Bose gas model,
this remarkable result was first implied by Mohling and Morita 
\cite{Mohling+:60} and explicitly obtained by Hohenberg and Martin
\cite{Hohenberg+:65} and by Popov \cite{Popov:72}. However, our
result is universally true for a generic system of scalar bosons in 
low temperature and low energy, free of the assumption of weak
coupling and low density. That implies that the result
(\ref{eq:gammaT}) also holds for liquid $^4$He. 

The low-energy excitation spectrum (\ref{eq:spectrum2}) is called
phonon spectrum and  has an important property: the excitation
frequency is equal to zero for zero momentum to all 
orders in interactions, which means that the spectrum does not exhibit
an energy gap.  This is known as the Hugenholtz-Pines theorem
\cite{Hugenholtz+:59} that was proved in perturbation field
theoretical analysis. It has played a crucial role in the early study of
superfluidity. Here we gain another understanding of this property.  It
is known from both microscopic theories
\cite{Gavoret+:64,Szepfalusy+:74} and  superfluid
hydrodynamics \cite{Hohenberg+:65} that,  if $T<T_c$,
the quasiparticle excitations have exactly the same
spectra as collective excitations (described here by the $\sigma$
field) in the long wavelength limit. Further,
the effective Lagrangian (\ref{eq:Leff2}) dictates that the Goldstone
mode $\phi$ and the collective mode $\sigma$ possess the same pole
structure up to all orders in all  interactions allowed by symmetries.
This statement is implied by Eq.~(\ref{eq:det}).
For the Goldstone theorem protects 
 all kinds of excitations in
low temperature and low momentum limit from having energy gap, their
spectrum must be of phonon type.
In this point of view, 
the Hugenholtz-Pines theorem  can be simply understood as the equivalent
statement of the {\em nonrelativistic} version
of  the Goldstone theorem
\cite{Lange:66,Katz+:66}  for the special case of a dilute Bose gas.

We conclude here that the effective field theory approach to the study
of collective excitations in BEC makes the role of the Bose--Einstein
condensate (or broken $U(1)$ symmetry) evident, and gives us a
model-independent effective action that can immediately predict
results for experiments.  We expect the approach to be very productive
and fundamental if one can conduct it for trapped alkali vapors that
are currently under extensive investigation.  Yet, the inhomogeneity
of the systems due to a trap potential shall alter our results.

\section*{Acknowledgement}

I am grateful to my advisor Steven Weinberg for tutoring
me about the effective field theory and the symmetry breaking,  for
very illuminating 
discussions that have  led essentially to the occurrence of the present
work, and for critically reading this manuscript. Also I would
like to thank Govindan Rajesh for helpful
conversations. 
This work is supported in part by NSF grant PHY-9511632 and the
Robert A. Welch Foundation.


\begin{thebibliography}{10}
\bibitem{Galilean:remark}
One can verify it by recalling the Galilean transformations: $t \rightarrow t$,
  ${\mbf x} \rightarrow {\mbf x} -{\mbf u} t$; ${1\over m} \nabla\phi
  \rightarrow {1\over m} \nabla\phi + {\mbf u}$, $i\partial_\tau\phi
  \rightarrow i\partial_\tau\phi + {\mbf u \cdot \nabla\phi} + {1\over 2}m
  u^2$; $\sigma \rightarrow \sigma$, $\nabla\sigma \rightarrow\nabla\sigma$,
  $i\partial_\tau\sigma \rightarrow i\partial_\tau\sigma +{\mbf u \cdot
  \nabla\sigma}$ (with $\tau=it$).
\bibitem{SigmaTerm:remark}
Alternatively, we should still include a term of $[i\partial_\tau\sigma
  -(1/m)\nabla\phi \cdot \nabla\sigma]^2$ in the most general action
  (\ref{eq:Ieff}). Accordingly, there would be a term of $(\partial_\tau
  \sigma)^2$ appearing in the quadratic effective action (\ref{eq:I^quad})
  below, which would lead to an additional $\omega^4$-term in evaluating the
  energy spectrum from Eq.~(\ref{eq:D(k)}). However, such a quartic term could
  be eliminated in comparison with quadratic terms since we are considering the
  lowest energy excitations. Thus, we would still have a linear spectrum and
  would derive the same power counting rule as (\ref{eq:nu2}). However, the
  rule would tell us that we do not need include any derivatives of $\sigma$
  from the beginning at all.
\end{thebibliography}
\end{document}